\documentclass[twocolumn,showpacs,preprintnumbers,amsmath,amssymb]{revtex4}

\usepackage{graphicx}
\usepackage{dcolumn}
\usepackage{bm}

\begin{document}

\preprint{.......}

\title{Locality hypothesis and the speed of light}

\author{G.Longhi}

\affiliation{Physics Department, University of
Florence}

\date{12 December 2005}

\begin{abstract}

The locality hypothesis is generally considered
necessary for the study of the kinematics of
non-inertial systems in Special Relativity. In this
paper we discuss this hypothesis, showing the necessity
of an improvement, in order to get a more clear
understanding of the various concepts involved, like
coordinate velocity and standard velocity of light.
Concrete examples are shown, where these concepts are
discussed.

\end{abstract}


\maketitle

\section{\label{S1} Introduction}

\bigskip

\noindent The study of the relativistic kinematics in
non-inertial reference frames has received recently
much attention \cite{RPau}, \cite{RLus1}, \cite{RRiz}.
Almost always the study of these frames is based on the
so called "locality hypothesis" which is the following
statement:

a) "an accelerated observer is \underline{equivalent}
to an infinite sequence of hypothetical inertial
observers along its world-line, each momentarily
co-moving with the accelerated observer,"\cite{RMas1}.
We can add the most obvious specification that the
observer should be a "point-like observer".

Another formulation of this hypothesis is given by:

b) "locally, neither gravity nor acceleration changes
the length of a standard rod or the rate of a standard
clock relative to a nearby freely falling standard rod
or a standard clock instantaneously co-moving with it".
And still "Stated another way, a local inertial
observer is \underline{equivalent} to a local co-moving
non-inertial observer in all matters having to do with
measurements of distance and time" \cite{RKla1}.

Let us quote even another formulation

c) "the speed of light, as measured \textit{locally} by
means of standard rods and clocks at rest with respect
to the non-inertial observer should be exactly
\underline{the same} as that observed in the local
inertial frame, the latter being c in both directions"
\cite{RTar1}.

Even if these formulations are all correct, some
ambiguity can arise if we do not clearly separate the
concepts of standard clocks and rods from that of
coordinate times and lengths. This because these last
quantities are in general quite different from the
standard ones.

For instance, in reference \cite{RKla1}, it is claimed
that the postulate, which says that the speed of light
is the same for all inertial observers and equal to c
(relativity postulate), could be violated in rotating
frames, if we maintain the locality hypothesis; but
again we observe that no clear distinction is made
there between the coordinates of the accelerated system
and that of the local inertial system.

Indeed, as will be seen in Section \ref{S6} , if we
maintain this distinction, no contradiction arises and
no violation of the relativity postulate or of the
locality hypothesis is required.

Now, in order to have a clear understanding of the
problem, it seems necessary to give in this context a
more precise notion of the aforementioned equivalence
or, which is the same, a clear relation between the
coordinates of a generally non-inertial system and the
local inertial coordinates.

In this work we want to recover a precise definition of
this equivalence and to show how it works in some
examples.

In Section \ref{S2} we will recall a well known fact
about the mathematics of Riemann geometry, namely that
is always possible to transform a given metric tensor
to a new one, which is locally Minkowskian and with
zero first derivatives. In this way it is possible to
define a local inertial frame, which can be used to
define the proper time and the proper lengths.

Of course this definition is well known, see for
instance the reference \cite{RMol1} or \cite{RWei1}.

With these results we can discuss the possibility of an
anisotropic propagation of light, and its relation with
the propagation as seen from the co-moving local
inertial frame.

In Section \ref{S3} we define some notation. In Section
\ref{S4} we study the example of a Galilean
transformation. This is the most simple example, but it
has already interesting characteristics regarding the
propagation of light.

In Section \ref{S5} we study the hyperbolic motion, and
in Section \ref{S6} the rotating disk.

In all these examples we look only at the propagation
of light and not of particles, this because the study
of the propagation of a material body would require the
complete solution of the geodesic equations. This could
be done, but it seems unnecessary for a clear
understanding of this matter.

In the examples that will be discussed, the existence
of a reference frame, inertial and with Minkowskian
coordinates, called laboratory frame, will be assumed.
In this way the accelerated systems will be easily
defined, but it must be understood that its existence
is not strictly necessary for the definition of the
local inertial frame.

In Section \ref{S7} is devoted to an application: the
solution of the so called Selleri's paradox.

We end in Section \ref{S8} with a concluding remark.

\bigskip
\bigskip

\section{\label{S2}The locality hypothesis}.

\bigskip

Let us start with a well known fact about Riemannian
geometry: given a metric tensor $g$ in a given
reference frame with coordinates $\{x^{\mu}\}$, free of
singularities and not degenerate, it is always
possible, in the neighborhood of a given point
$P_{\circ}$, to find a transformation to a new set of
coordinates $\{x^{\prime \mu}\}$, such that $g$ will
transform in a Minkowski metric in the point
$P_{\circ}$, and such that its first derivatives in
this point be zero (normal coordinates). In general it
is not possible to require the vanishing of the higher
derivatives of $g$.

This transformation is determined up to a Lorentz
transformation \cite{RWei1}, \cite{RMol1}.

Following \cite{RMol1}, but see also \cite{RWei1} and
\cite{RWhe}, this transformation can be written

\begin{equation}\label{E2.1}
x^{\prime \mu} = b^{\mu}_{\alpha}(x^{\alpha} -
x^{\alpha}_{\circ}) + \frac{1}{2}
b^{\mu}_{\nu}\Gamma^{\nu}_{\circ\alpha\beta}
(x^{\alpha} - x^{\alpha}_{\circ})(x^{\beta} -
x^{\beta}_{\circ}),
\end{equation}

\noindent  where $\{x^{\mu}_{\circ}\}$ are the
coordinates of $P_{\circ}$, which is the origin of the
new coordinates $\{x^{\prime\mu}\}$, and
$\Gamma_{\circ}\equiv\Gamma(P_{\circ})$ are the
connection coefficients of the metric tensor $g$ at the
point $P_{\circ}$. We will always use coordinate basis
for the vector fields, so that the connection
coefficients $\Gamma$ are given by

\begin{equation}\label{E2.2}
\Gamma^{\mu}_{\alpha\beta} = \frac{1}{2} g^{\mu\nu}
(\frac{\partial g_{\nu\alpha}}{\partial x^{\beta}} +
\frac{\partial g_{\nu\beta}}{\partial x^{\alpha}} -
\frac{\partial g_{\alpha\beta}}{\partial x^{\nu}}).
\end{equation}

The matrix $b$ must be chosen such that

\begin{equation}\label{E2.3}
\eta_{\alpha\beta}b^{\alpha}_{\mu}b^{\beta}_{\nu} =
g_{\mu\nu}(P_{\circ}),
\end{equation}

\noindent where $\eta$ is the metric of Minkowski, with
signature $(-,+++)$.

It can be shown that the matrix transformation
(\ref{E2.3}) is always possible \cite{RWei2}, and we
will see it explicitly in the examples of the following
Sections.

Once $g$ is reduced to $\eta$ in $P_{\circ}$ we have
clearly still the freedom to perform an arbitrary
Lorentz transformation. But this can be uniquely
determined if we require in addition that the
coordinates axis be tangent to a given tetrad of
vectors in $P_{\circ}$. In particular the coordinate
axis of the time $x^{\prime 0} = ct^{\prime}$ will be
chosen tangent to the axis of $x^0 = ct$, in this way
the frame defined by the new coordinates will be a
system co-moving with $P_{\circ}$.

In this sense we may say that the transformation
$\{x^{\mu}\}\rightarrow \{x^{\prime\mu}\}$ is unique.

Let us now look at the geodesic equations. These are

\begin{equation}\label{E2.4}
\frac{d^2 x^{\mu}}{d \lambda^2} +
\Gamma^{\mu}_{\alpha\beta} \frac{d x^{\alpha}}{d
\lambda}\frac{d x^{\beta}}{d \lambda} = 0,
\end{equation}

\noindent and, in the case of null geodesics, we must
add

\begin{equation}\label{E2.5}
g_{\mu\nu}\frac{d x^{\mu}}{d \lambda}\frac{d x^{\nu}}{d
\lambda} = 0,
\end{equation}

\noindent with $\lambda$ some parameter.

These are the geodesic equations in the $\{x^{\mu}\}$
coordinates. But in terms of the $\{x^{\prime\mu}\}$
coordinates \underline{in the point $P_{\circ}$} these
equations become

\begin{equation}\label{E2.6}
\begin{cases}
& \frac{d^2 x^{\prime\mu}}{d \lambda^2}|_{P_{\circ}} = 0,\\
& \eta_{\mu\nu}\frac{d x^{\prime\mu}}{d \lambda}\frac{d
x^{\prime\nu}}{d \lambda}|_{P_{\circ}} = 0,\\
\end{cases}
\end{equation}

\noindent since in $P_{\circ}$ the connection
coefficients $\Gamma_{\circ}$, being proportional to
the first derivatives of the metric tensor, are zero.

So the geodesic equations in the new coordinates give
rise to a light propagation isotropic and along
trajectories, which are rectilinear in a second order
neighborhood of the point $P_{\circ}$.

More exactly, the word line with $\lambda$ as a
parameter will be of the form

\begin{equation}\label{E2.7}
x^{\prime\mu} = a^{\mu} + b^{\mu} \lambda + {\rm
O}(\lambda^{^3}),
\end{equation}

\noindent since the second order term is zero due to
equation (\ref{E2.6}). This equation determines the
degree of approximation of the local inertial frame to
the system S.

On the other hand, the new metric tensor
$g^{\prime}_{\mu\nu}dx^{\prime \mu}dx^{\prime\nu}$ will
be inertial only up to first order in
$x^{\mu}-x^{\mu}_{\circ}$, as seen from (\ref{E2.1}).

If we now apply the previous result to the relativistic
kinematics of a non-inertial frame with a generic
metric tensor $g$, we see that it is quite natural to
define the aforementioned equivalence in terms of the
transformation $\{x^{\mu}\}\rightarrow
\{x^{\prime\mu}\}$. So we can change the locality
hypothesis (a) to the following form:

"an accelerated observer is equivalent, by means of a
sequence of \underline{coordinate}
\underline{transformation} (\ref{E2.1}), to an infinite
sequence of inertial observers along its world-line,
each momentarily co-moving with the accelerated
observer,".

If we agree that proper times and distances are the
times and distances measured by standard clocks and
rods, that is clocks and rods at rest in a given place
and at a given time on the accelerated frame, we can
interpret this hypothesis by saying: the measure of
proper times and distances in a given accelerated
system are given by the times and the distances as
measured by using the coordinates of the local inertial
frame, which is co-moving with the accelerated system,
at the same place and time.

We will see that ambiguities in the application of the
principles of relativity, in situations in which the
propagation of light is anisotropic in one reference
frame and isotropic in another, will be in principle
solved.

This will be essentially achieved by observing that two
observers, practically in the same place and at the
same time, one performing measures of time and
distances in terms of the $\{x^{\mu}\}$ coordinates and
the other in terms of the $\{x^{\prime\mu}\}$, will see
a \underline{different} propagation of light.

We will see in detail this fact in the examples of the
following Sections, but we may easily understand  how
it may happen by considering a 2-dimensional system
described by a set of coordinates $\{cT,X\}$ with
Minkowskian metric

\begin{equation}\label{E2.8}
\eta = \eta_{\mu\nu}dX^{\mu}dX^{\nu} = -c^2dT^2 + dX^2.
\end{equation}

The light-cone equation with the vertex in the origin
is, of course

\begin{equation}\label{E2.9}
-c^2 T^2 + X^2 = 0.
\end{equation}

If we perform a linear transformation to a new set of
coordinates $\{cT, X\} \rightarrow \{ct, x\}$, which is
\underline{not} a Lorentz transformation, and if we
choose this transformation such that to give a mixing
of the spatial and the temporal coordinates, the new
metric will be \underline{not time-orthogonal}.

Of course, any transformation regular enough can be
Taylor expanded, and can be considered a linear
transformation in the neighborhood of a given point in
some approximation.

The result of this transformation will be a
\underline{deformation} of the light cone in such a way
that the speed of light in the forward direction is,
say, higher than c and in the backward direction less
than c. If for instance this linear transformation is
(Galilean transformation)

\begin{eqnarray}\label{E2.10}
\begin{cases}
t &= T,\\
x &= X - v T,\\
\end{cases}
\end{eqnarray}

\noindent the light-cone equation becomes

\begin{equation}\label{E2.11}
-c^2(1 - \beta^2) t^2 + 2 v x t + x^2 = 0,
\end{equation}

\noindent where $\beta = v/c$. The light-cone which was
symmetric with respect to the time axis, is no more
symmetric, and the speed of light is c+v in the
positive direction of the x axis and -c+v in the other.

This situation is analogous to the propagation of the
light from a star, coming from a visual position near
to the sun. Suppose that this light is observed from a
local inertial frame and from the astronomical
non-inertial system. In the first case we have a
rectilinear propagation with velocity c, in the second
case we have a deflection \cite{RWei3}.

In this example we have that the synchronization in the
accelerated frame is determined by the laboratory
frame, where the standard synchronization (Einstein
synchronization) is supposed to hold. This will be true
in all the examples considered in the following.

The composition of velocities that follows from this
transformation is the Galilean composition law. Since
the system is 1-dimensional there is no room for an
explanation in terms of a normal velocity, as done in
reference \cite{RPet}, on the contrary, this example
shows that the explanation given in that reference
cannot be maintained. As any other transformation it
has its own composition law. We will study this
transformation in more details below.

Several topics, which could be studied using the
transformation (\ref{E2.1}), will not be discussed. For
instance, the spatial geometry, that is the surface of
constant proper time, or the limit of validity of the
locality hypothesis, when interpreted in the large, as
done in \cite{RMas1}, \cite{RMas2} will not be studied.

\bigskip
\bigskip

\section{\label{S3}The co-moving local inertial frame.}

\bigskip

In the examples of the following Sections we will use
the following notations: the coordinates of a inertial
\underline{laboratory frame} $\mathcal L$ will be
denoted $\{X^{\mu}\}$, with $X^0 = cT$. It is tacitly
supposed that in $\mathcal L$ the clocks are
synchronized according to the Einstein rule
\cite{RSte}. The metric tensor will be

\begin{equation}\label{E3.1}
\eta_{\mu\nu}dX^{\mu}dX^{\nu} = -c^2dT^2 + dX^2 + dY^2
+ dZ^2,
\end{equation}

\noindent and an accelerated system $\mathcal S$ will
be defined with respect to this system.

We must stress that the existence of the frame
$\mathcal L$ is not necessary, but it is useful for
deducing the metric of $\mathcal S$.

The coordinates of the \underline{accelerated system}
$\mathcal S$ will be denoted $\{x^{\mu}\}$, with $x^0 =
ct$.

The synchronization of clocks in $\mathcal S$ is not
arbitrary, but it is borrowed from that of $\mathcal
L$. Usually the synchronization is characterized by the
quantity $\epsilon$ \cite{RHav} \cite{RCap}, see the
Appendix, which, in the case of the Einstein
synchronization, is $1/2$ . In $\mathcal S$ $\epsilon$
can be different from $1/2$.

Finally, for the co-moving \underline{local inertial
system} $\mathcal L^{\prime}$, we will use the
coordinates $\{x^{\prime\mu}\}$, with origin in the
generic point $P_{\circ}\equiv \{x^{\mu}_{\circ}\}$.
This system will be determined by the transformation
$\{x^{\mu}\} \rightarrow \{x^{\prime\mu}\}$ and it will
be connected to the system $\mathcal L$ by means of a
inhomogeneous Lorentz transformation.

Now, the value of $\epsilon = 1/2$ is conserved under
Lorentz transformation. This means that, if we define
the $\{x^{\prime\mu}\}$ in terms of the $\{X^{\mu}\}$,
since these coordinates are connected by a Lorentz
transformation, the synchronization in $\mathcal
L^{\prime}$ will be still with $\epsilon = 1/2$, and
this will be true for each local inertial frame.

An observation which is almost obvious, is the
following: if we want to consider the transformation
from the system $\mathcal L$ to the system $\mathcal S$
as a physical transformation, and not merely as a
change of notations, we must suppose that
\underline{experimental procedures} for the measure of
the $\mathcal S$ coordinates be provided. So it must be
possible the measure of the speed $\frac{d {\vec x}}{d
t}$, which will be called \underline{coordinate speed}
\cite{RPet},\cite{RWuc}. This seems obvious, but it can
be practically difficult in some situation. For
instance, in the case of a rotating disk it can be not
easy to measure the coordinate time $t$, by an observer
rotating with the disk, since it is quite different
from the proper time.

The $\mathcal S$ coordinates will be called the
\underline{coordinate} \underline{time} and
\underline{space} respectively, while the coordinates
$\{x^{\prime\mu}\}$ of $\mathcal L^{\prime}$ are by
definition the \underline{proper time} ($x^{\prime
0}$), that is the time measured with a standard clock
in $P_{\circ}$, and the \underline{proper space}
(${\vec x}^{\prime}$), measured with a standard rod in
the same place.

The speed $\frac{d {\vec x^{\prime}}}{d t^{\prime}}$
will be called the \underline{standard speed}. For what
we have said before the standard speed of light will be
always isotropic and of value c, while the coordinate
speed can be anisotropic and different from c. It is
the standard speed that is measured locally by standard
clocks and rods.

\bigskip
\bigskip

\section{\label{S4} The Galilean transformation}

\bigskip

Let us consider again the Galilean transformation from
our  laboratory frame $\mathcal L$ to $\mathcal S$,
defined as in \cite{RMolx}

\begin{equation}\label{E4.1}
\begin{cases}
c T \rightarrow c t = c T,\\
X \rightarrow x = X - vT,\\
Y = y,\quad Z = z.\\
\end{cases}
\end{equation}

In this Section the coordinates $Y, y$ and $Z, z$ will
be omitted.

The metric tensor in the new coordinates is

\begin{equation}\label{E4.2}
g = -c^2(1-\beta^2) dt^2 + dx^2 + 2\beta dx c dt,
\end{equation}

\noindent where $\beta = v/c$.

Since the metric components are constant, the
connection coefficients $\Gamma$ will be zero, as a
consequence the geodesic equation in $\mathcal S$ will
be as

\begin{equation}\label{E4.3}
\frac{d^2 x^{\mu}}{d \lambda^2} = 0.
\end{equation}

Nevertheless, as we have seen, the propagation of light
is \underline{anisotropic}. Indeed, for the propagation
of light we have, from (\ref{E4.2}),

\begin{equation}\label{E4.4}
-c^2(1-\beta^2) dt^2 + dx^2 + 2\beta dx c dt = 0,
\end{equation}

\noindent or

\begin{equation}\label{E4.5}
-c^2(1-\beta^2) + \dot{x}^2 + 2\beta c \dot{x} =0,
\end{equation}

\noindent where $\dot{x}= \frac{dx}{dt}$, from which

\begin{equation}\label{E4.6}
\dot{x} = \pm c - v.
\end{equation}

The synchronization parameter is $\epsilon =
\frac{1}{2} (1 \pm \beta)$.

Let us consider an event $P_{\circ}$ fixed in the
$\mathcal S$ system , with coordinates $(t_{\circ},
x_{\circ})$. In the $\mathcal L$ system it will
describe a world-line

\begin{equation}\label{E4.7}
X = x_{\circ} + v t,\quad {\rm and}\quad T = t.
\end{equation}

The vector $e_{\circ}$, defined as

\begin{equation}\label{E4.8}
e_{\circ} = \gamma (\frac{\partial}{\partial X^{\circ}}
+ \beta\frac{\partial}{\partial X}),\quad {\rm
with}\quad e_{\circ}^2 = -1,
\end{equation}

\noindent where $\gamma = (1-\beta^2)^{-\frac{1}{2}}$,
is tangent to the world line, and the vector $e_1$

\begin{equation}\label{E4.9}
e_1 = \gamma (\beta\frac{\partial}{\partial X^{\circ}}
+ \frac{\partial}{\partial X}),\quad {\rm with}\quad
e_1^2 = 1,\quad (e_1\ |\ e_{\circ}) = 0,
\end{equation}

\noindent where $(\ |\ )$ is the metric scalar product
and $X^{\circ} = c T$, is the vector orthogonal to
$e_{\circ}$.

Let us now determine the transformation $x^{\mu}
\rightarrow x^{\prime\mu}$, using equations
(\ref{E2.1}) and (\ref{E2.3}). Since the $\Gamma$
coefficients are zero we need only to find the matrix
$b^{\mu}\,_{\nu}$. This is easily found:

\begin{equation}\label{E4.10}
\parallel b^{\mu}\,_{\nu}\parallel =
{\setlength\arraycolsep{5pt} \left(\begin{array}{rr}
\gamma(1-\beta^2) &   -\gamma\beta\\
0\quad\quad\quad  &    \gamma\\
\end{array}\right).}
\end{equation}

With this matrix we get for the coordinates
$x^{\prime\mu}$ (with $x^{\prime\circ} = c t^{\prime}$,
$x^{\prime 1} = x^{\prime}$, $x^{\circ} = c t$,
$x\equiv x^1$)

\begin{eqnarray}\label{E4.11}
\begin{cases}
x^{\prime\circ} &= \gamma[(1-\beta^2) (x^{\circ} -
x^{\circ}_{\circ}) - \beta (x - x_{\circ})],\\
x^{\prime} &= \gamma(x - x_{\circ}).\\
\end{cases}
\end{eqnarray}

It is easily verified that the new metric tensor
$g^{\prime}$ is, as expected,

\begin{equation}\label{E4.12}
g' = -c^2 dt^{\prime 2} + dx^{\prime 2},
\end{equation}

\noindent and, using both transformations (\ref{E4.1})
and (\ref{E4.11}), we may verify that the coordinate
axis are tangent to the vectors $e_{\circ}$ and $e_1$:

\begin{equation}\label{E4.13}
\begin{cases}
\frac{\partial\ }{\partial x^{\prime\circ}} &= e_{\circ},\\
\frac{\partial\ }{\partial x^{\prime}} &= e_1.\\
\end{cases}
\end{equation}

With these two conditions satisfied the transformation
(\ref{E4.11}) is unique. The transformation
(\ref{E4.11}) defines a co-moving, local inertial frame
in the neighborhood of the point $P_{\circ}$.

Finally we may determine the relation between the
velocities. If $u$ is a velocity as measured in
$\mathcal S$ and $u^{\prime}$ is the corresponding
velocity in $\mathcal L^{\prime}$, we may easily find
their relation from equations (\ref{E4.11})

\begin{equation}\label{E4.14}
u^{\prime} = \frac{u}{(1-\beta^2) - \beta\frac{u}{c}}.
\end{equation}

The velocity $u$, in the case of the velocity of the
light, is given by (\ref{E4.6}), so we get

\begin{equation}\label{E4.15}
u^{\prime} = \frac{\pm c(1 \mp\beta)}
{1-\beta^2\mp\beta+\beta^2} = \pm c,
\end{equation}

\noindent so the anisotropy of the propagation of light
disappears in the local inertial frame.

It general it is also possible to define an "extended"
frame, whose coordinates, up to second order in
$x^{\mu} - x^{\mu}_{\circ}$, be identical with the
$\{x^{\prime\mu}\}$ of equation (\ref{E2.1}). In the
present case they are linear, so no approximation is
needed and the coordinates given by equation
(\ref{E4.11}) are the coordinates of the extended frame
too. In any case, this extended frame could approximate
the accelerated one only in a neighborhood of the point
of interest.

\bigskip
\bigskip

\section{\label{S5} The hyperbolic motion.}

\bigskip

Following the same steps of the previous example, we
may study the hyperbolic motion in the plane $(cT,\,X)$
\cite{RMol2}.

Let us consider a congruence of time-like world lines
given by

\begin{equation}\label{E5.1}
X = X_{\circ}+{\frac{c^2}{g}}(\sqrt{1+(\frac{g
T}{c})^2} -1),
\end{equation}

\noindent where $g$ is a given acceleration and
$X_{\circ}$ is the initial value of $X$. Varying
$X_{\circ}$ we get the various world lines of the
congruence. In the $(T,X)$ plane, these lines are, as
well known, hyperbolas.

In the limit $g T \ll c$ we get simply

\begin{equation}\label{E5.2}
X \simeq X_{\circ} + \frac{1}{2} g T^2.
\end{equation}

We now consider a new set of coordinates obtained with
the substitution $X_{\circ}\rightarrow x$ and with $T
=t$:

\begin{eqnarray}\label{E5.3}
\begin{cases}
x &= X -\frac{c^2}{g}(\sqrt{1 + (\frac{g T}{c})^2}-1),\\
t &= T.\\
\end{cases}
\end{eqnarray}

The new set of coordinates $(t,x)$ defines the
accelerated system $S$. We could have expressed the
time $T$ in terms of the proper time, as done in
\cite{RWei4} and \cite{RMas3}, but we prefer to
maintain the Cartesian coordinates.

The initial Minkowskian metric $\eta$ is transformed to

\begin{equation}\label{E5.4}
g = g_{\mu\nu}dx^{\mu}dx^{\nu} = - c^2(1-\beta^2) dt^2
+ dx^2 + 2c\beta dx dt,
\end{equation}

\noindent where

\begin{equation}\label{E5.5}
\beta =(\frac{g
t}{c})\frac{1}{\sqrt{1+(\frac{gt}{c})^2}},
\end{equation}

\noindent which is not \underline{time-orthogonal}.

The vector tangent to these world-lines is

\begin{equation}\label{E5.6}
e_{\circ} = \gamma (\frac{\partial}{\partial X^{\circ}}
+ \beta\frac{\partial}{\partial X}), \quad  \quad
(e_{\circ})^2 = -1,
\end{equation}

\noindent and the vector orthogonal to $e_{\circ}$ is

\begin{equation}\label{E5.7}
e_1 = \gamma (\beta\frac{\partial}{\partial X^{\circ}}
+ \frac{\partial}{\partial X}), \quad  (e_1)^2 = +1,
\quad (e_1\ |\ e_{\circ}) = 0.
\end{equation}

In these equations

\begin{equation}\label{E5.8}
\gamma = \sqrt{1 + (\frac{g t}{c})^2}= (1
-\beta^2)^{-1/2}.
\end{equation}

Even in this case we have an anisotropic propagation of
light. Indeed, from equation (\ref{E5.4}) for the null
geodesics

\begin{equation}\label{E5.9}
-c^2 (1 - \beta^2) +(\frac{dx}{dt})^2 + 2 \beta c
\frac{dx}{dt} = 0,
\end{equation}

\noindent we get

\begin{equation}\label{E5.10}
\frac{dx}{dt} = \pm c(1 \mp \beta).
\end{equation}

The synchronization parameter is
$\epsilon=\frac{1}{2}(1\pm\beta)$.

Now we want to determine the transformation ${x^{\mu}}
\rightarrow {x^{\prime\mu}}$, in the neighborhood of a
generic point $P_{\circ} \equiv (t_{\circ},
x_{\circ})$. Since

\begin{equation}\label{E5.12}
\parallel g_{\mu\nu}\parallel =
{\setlength\arraycolsep{5pt} \left(\begin{array}{rr}
   - (1-\beta^2)      &   \beta\\
\beta\quad\quad\quad  &    1   \\
\end{array}\right),}
\end{equation}

\noindent from (\ref{E2.2}) we get that the only
connection coefficient different from zero is

\begin{equation}\label{E5.13}
\Gamma^1_{00} = \frac{g}{c^2}(1-\beta^2)^{3/2}.
\end{equation}

The matrix $\parallel b^{\mu}_{\nu}\parallel$ is easily
calculated

\begin{equation}\label{E5.14}
\parallel b^{\mu}\,_{\nu}\parallel =
{\setlength\arraycolsep{5pt} \left(\begin{array}{rr}
\gamma_{\circ}(1-\beta_{\circ}^2)&-\gamma_{\circ}\beta_{\circ}\\
     0\quad\quad\quad            &    \gamma_{\circ}\\
\end{array}\right),}
\end{equation}

\noindent where $\gamma_{\circ}$ and $\beta_{\circ}$
are $\gamma$ and $\beta$ in the point $P_{\circ}$.

Substituting this matrix and $\Gamma$ in equation
(\ref{E2.1}), we finally get (see also \cite{RMas3}
where a transformation of the same kind is determined)

\begin{eqnarray}\label{E5.15}
\begin{cases}
x^{\prime\circ}
&=\gamma_{\circ}\{(1-\beta_{\circ}^2)\eta^{\circ} -
\beta_{\circ}\eta^1-\frac{g\beta_{\circ}}
{2c^2\gamma_{\circ}^3}(\eta^{\circ})^2\},\\
x^{\prime} &= \gamma_{\circ}\{\eta^1+
\frac{g}{2c^2\gamma_{\circ}^3}(\eta^{\circ})^2\},\\
\end{cases}
\end{eqnarray}

\noindent where $x^{\prime} \equiv x^{\prime 1}$,
$x^{\circ} = c t$ and $\eta^{\mu} = x^{\mu} -
x_{\circ}^{\mu}$.

This is the transformation we were looking for.

We could as well have found the transformation
${X^{\mu}} \rightarrow {x^{\prime\mu}}$, using the
relation (\ref{E5.3}). In this way it is possible to
verify, with some work, that the coordinate axis of the
new variables are tangent to the vectors $e_{\circ}$
and $e_1$ in $P_{\circ}$:

\begin{equation}\label{E5.16}
\begin{cases}
\frac{\partial\ }{\partial x^{\prime\circ}} &= e_{\circ},\\
\frac{\partial\ }{\partial x^{\prime}} &= e_1.\\
\end{cases}
\end{equation}

We may too verify that the new metric tensor $g'$ is
Minkowskian

\begin{equation}\label{E5.17}
g'_{\mu\nu}(x')dx'^{\mu}dx'^{\nu} = - c^2 dt'^2 +
dx'^2,
\end{equation}

\noindent which is true up to terms of order
$O((x')^2)$, or even of order $O((\eta)^2)$, since the
relation between the $x'$ and the $\eta$ is
homogeneous.

More precisely we can write

\begin{equation}\label{E5.17bis}
\begin{cases}
&g^{\prime}_{\mu\nu}(x^{\prime})dx^{\prime\mu}dx^{\prime\nu}
= [\eta_{\mu\nu} +
O((x')^2)]dx^{\prime\mu}dx^{\prime\nu} =\\
&= [g_{\mu\nu}(x) + O((\eta^{\mu})^2)]dx^{\mu}dx^{\nu}.
\end{cases}
\end{equation}

\noindent where $g$ is the metric (\ref{E5.4}). This
means that the first order derivatives in $P_{\circ}$
of $g'$ are zero, as expected.

With the two conditions (\ref{E5.16}) and (\ref{E5.17})
the transformation (\ref{E5.15}) is unique.

From (\ref{E5.15}) we may get the relation between the
velocities as seen from the two frames in the point
$P_{\circ}$

\begin{equation}\label{E5.18}
\frac{d x^{\prime}}{d t^{\prime}} = \frac{\frac{d x}{d
t}} {1-\frac{\beta_{\circ}}{c}\frac{d x}{d t} -
\beta_{\circ}^2}
\end{equation}

We have seen before in equation (\ref{E5.10}) that the
coordinate velocity $\frac{dx}{dt}$, measured in
$\mathcal S$, is anisotropic. Substituting in the
previous equation we obtain for the standard velocity
$\frac{d x^{\prime}}{d t^{\prime}}$, measured in
${\mathcal L}^{\prime}$ in the point $P_{\circ}$

\begin{equation}\label{E5.19}
\frac{d x^{\prime}}{d t^{\prime}} = \pm c,
\end{equation}

\noindent and there is \underline{no} more anisotropy.

\bigskip
\bigskip

\section{\label{S6} The rotating platform}

\bigskip

The rotating platform is an example so much studied
that we could hardly do justice to all contributors. We
may refer for an extended set of references in the book
edited by G.Rizzi and M.T.Ruggiero \cite{RRiz}, where a
consistent amount of bibliography can be found.

In any case we can quote here a set of articles on this
subject, which have a point of view similar to that
adopted here.

In \cite{RMas2} an analysis similar to our, with the
end of verifying the limit of validity of the locality
hypothesis, is performed for the rotating disk.

In \cite{RPer} the synchronization of clocks on a
rotating disk is discussed. On the other hand a
discussion of all the aspect of the kinematics of the
rotating disk can be found in \cite{RCan}, \cite{RGro}
and \cite{RStr}.

We do not discuss here the problem of the
desynchronization of clocks, or Sagnac effect, for a
round trip along a circumference of the disk. This
because we directed our attention on the formulation of
the locality hypothesis. This points are discussed in a
lot of papers. Quoting some of them we have for
instance \cite{RAna}, \cite{RSel}, \cite{RTar1},
\cite{RBel} and \cite{RKla1}.

In order to avoid complications with the rigidity
constraint, we will not study really a rotating disk,
but rather we will consider a model defined by a
congruence of time-like helices, given in the reference
frame $\mathcal L$ by the equations

\begin{equation}\label{E6.1}
\begin{cases}
T &= t,\\
X &= x\cos(\omega t)- y\sin(\omega t),\\
Y &= x\sin(\omega t) + y\cos(\omega t),\\
\end{cases}
\end{equation}

\noindent where $T, X, Y$ are the cartesian coordinates
in $\mathcal L$, and $x, y$ are the values of $X$ and
$Y$ when the parameter $t$ is zero. Nevertheless, for
sake of simplicity and intuition, we will continue to
speak of a disk, of the rim of this disk and so on.

The helices of the congruence are obtained by varying
the values of $x$ and $y$.

From now on we will omit the $z$ coordinate.

The vector $e_{\circ}$ tangent to the helix is given by

\begin{equation}\label{E6.2}
e_0 \propto\dot{X}^{\mu}\frac{\partial}{\partial
X^{\mu}}.
\end{equation}

If we normalize it to $-1$ and use polar coordinates

\begin{equation}\label{E6.2bis}
X = R\cos{(\Phi)},\quad Y = R\sin{(\Phi)},
\end{equation}

\noindent we get

\begin{equation}\label{E6.3}
\begin{cases}
e_0 &= \gamma(\frac{\partial}{\partial X^{\circ}} +
\frac{\omega}{c}\frac{\partial}{\partial \Phi}),\\
e_0^2 &= -1,\\
\end{cases}
\end{equation}

\noindent where $\gamma=1/\sqrt{1-\beta^2}$, and
$\beta=\omega r/c$.

We have two other vectors orthogonal each other and to
$e_0$,

\begin{equation}\label{E6.4}
\begin{cases}
e_1 &= \frac{\partial}{\partial R},\\
e_2 &= \gamma(\beta\frac{\partial}{\partial X^{\circ}}
+
\frac{1}{R}\frac{\partial}{\partial \Phi}), \\
\end{cases}
\end{equation}

\noindent with

\begin{equation}\label{E6.5}
\begin{cases}
e_1^2 = +1,\quad e_2^2 = +1,\quad (e_1| e_2) = 0,\\
\nonumber (e_1| e_0) = (e_2| e_0) = 0.\\
\end{cases}
\nonumber\end{equation}

The transformation to $\mathcal S$ is given by

\begin{equation}\label{E6.7}
r = R,\quad\quad  \phi = \Phi - \omega T,\quad t = T,
\end{equation}

\noindent where

\begin{equation}\label{E6.7bis}
x = r\cos{(\phi)},\quad y = r\sin{(\phi)},
\end{equation}

In cartesian coordinates the transformation
$(X,Y,T)\rightarrow(x,y,t)$ is

\begin{equation}\label{E6.8}
\begin{cases}
x = X\cos{(\omega T)} + Y\sin{(\omega T)},\\
y = -X\sin{(\omega T)} +Y\cos{(\omega T)},\\
\end{cases}
\end{equation}

\noindent and $t =T$.

The metric tensor in the frame $\mathcal S$ becomes

\begin{eqnarray}\label{E6.9}
&g = g_{\mu\nu}dx^{\mu}dx^{\nu} =\\
&-c^2(1-\beta^2)dt^2
+ dx^2 + dy^2 + 2\omega(-ydx + xdy) dt,\\
\nonumber\end{eqnarray}

\noindent which is not \underline{time-orthogonal}.

In polar coordinates it is given by

\begin{equation}\label{E6.10}
g = -c^2(1-\beta^2)dt^2 + dr^2 + r^2d\phi^2 + 2\omega
r^2 dt d\phi.
\end{equation}

The coordinate speed of light can be obtained from the
equation $g = 0$. For the radial speed in $P_{\circ} $
we get

\begin{equation}\label{E6.12}
\dot{r}_{\circ} = \pm c\sqrt{1-\beta_{\circ}^2},\quad
\dot{\phi}_{\circ} = 0,
\end{equation}

\noindent and for the tangential velocity

\begin{equation}\label{E6.13}
(r\dot{\phi})_{\circ} = \pm c(1\mp\beta_{\circ}),\quad
\dot{r}_{\circ} = 0,
\end{equation}

\noindent where $\beta_{\circ}=\omega r_{\circ}/c$.

We see that there is anisotropy for the tangential
propagation and not for the radial one. On the other
hand, in the radial case, the value of the coordinate
speed is different from c.

The synchronization parameter is given by
$\epsilon=\frac{1}{2}(1\pm\beta_{\circ})$.

In order to find the coordinates of $\mathcal
L^{\prime}$ we need the connection coefficients. The
only  connection coefficients different from zero are:

\begin{equation}\label{E6.14}
\begin{cases}
\Gamma^1_{00} = - \frac{\omega^2 x}{c^2},\quad
&\Gamma^1_{02} = - \frac{\omega}{c},\\
\Gamma^2_{00} = - \frac{\omega^2 y}{c^2},\quad
&\Gamma^2_{01} = + \frac{\omega}{c}.\\
\end{cases}
\end{equation}

The matrix $\parallel b^{\mu}_{\nu}\parallel$ of
equation (\ref{E2.1}) is easily found

\begin{equation}\label{E6.15}
\parallel b^{\mu}_{\nu}\parallel = \Lambda\textrm{R},
\end{equation}

\noindent where

\begin{equation}\label{E6.16}
\parallel \Lambda^{\mu}\,_{\nu}\parallel =
{\setlength\arraycolsep{5pt} \left(\begin{array}{rrr}
\gamma_{\circ}(1-\beta_{\circ}^2) & 0 &
-\gamma_{\circ}\beta_{\circ}\\
0\quad\quad & 1 & 0\\
0\quad\quad & 0 & \gamma_{\circ}\\
\end{array}\right).}
\end{equation}

where $\mu, \nu = 0,1,2$, and

\begin{equation}\label{E6.17}
\parallel R^{\mu}\,_{\nu}\parallel =
{\setlength\arraycolsep{5pt} \left(\begin{array}{rrr}
1 & 0 & 0\\
0 & \cos{(\phi_{\circ})} & \sin{(\phi_{\circ})}\\
0 & -\sin{(\phi_{\circ})} & \cos{(\phi_{\circ})}\\
\end{array}\right),}
\end{equation}

\noindent where $r_{\circ}$, $\phi_{\circ}$ and
$x^{\circ}_{\circ} $, $x^1_{\circ}$ and $x^2_{\circ}$
are the coordinates of the point $P_{\circ}$.

With these matrices we have, in $P_{\circ}$

\begin{equation}\label{E6.17bis}
((b^{-1})^T g (b^{-1})) = \eta.
\end{equation}

Substituting in equation (\ref{E2.1}) we get the
transformation $\{x^{\mu}\} \rightarrow
\{x^{\prime\mu}\}$ (see \cite{RMas2} for a similar
transformation)

\begin{equation}\label{E6.18}
\begin{cases}
x^{\prime\circ} &=
\gamma_{\circ}\{(1-\beta_{\circ}^2)\eta^{\circ}
+\beta_{\circ}[\sin{\phi_{\circ}}\eta^1 -
\cos{\phi_{\circ}}\eta^2] -\\
 &-\beta_{\circ}[\cos{\phi_{\circ}}\eta^1+
\sin{\phi_{\circ}}\eta^2]\frac{\omega}{c}\eta^{\circ}\},\\
\end{cases}
\end{equation}

\begin{equation}\label{E6.19}
\begin{cases}
x^{\prime 1} = &\cos{\phi_{\circ}}\eta^1 +
\sin{\phi_{\circ}}\eta^2 +\\
& +\frac{\omega}{c}(\sin{\phi_{\circ}} \eta^1 -
\cos{\phi_{\circ}}\eta^2)\eta^{\circ}-
\frac{\omega^2}{2c^2}r_{\circ}(\eta^{\circ})^2,\\
\end{cases}
\end{equation}

\begin{equation}\label{E6.20}
\begin{cases}
x^{\prime 2} =
&\gamma_{\circ}\{-\sin{\phi_{\circ}}\eta^1+
\cos{\phi_{\circ}}\eta^2+\\
& + \frac{\omega}{c}[\cos{\phi_{\circ}}\eta^1+
\sin{\phi_{\circ}}\eta^2]\eta^{\circ}\},\\
\end{cases}
\end{equation}

\noindent where, as before, $\eta^{\mu} = x^{\mu} -
x^{\mu}_{\circ}$, $x^{\circ} = ct, x^1 = x, x^2 = y $
etc.

Using the previous transformation we must remember that

\begin{equation}\label{E6.21}
\begin{cases}
x^1_{\circ}\cos{\phi_{\circ}} +
x^2_{\circ}\sin{\phi_{\circ}} &= r_{\circ},\\
x^1_{\circ}\sin{\phi_{\circ}} -
x^2_{\circ}\cos{\phi_{\circ}} &= 0.\\
\end{cases}
\end{equation}

The new metric tensor is Minkowskian, and

\begin{equation}\label{E6.22}
\begin{cases}
&g^{\prime}_{\mu\nu}(x')dx^{\prime\mu}dx^{\prime\nu} =
-(dx^{\prime\circ})^2 + (dx^{\prime 1})^2 + (dx^{\prime
2})^2  =\\
&-c^2(1-\beta^2)dt^2 + dx^2 + dy^2 +
2\omega(-ydx+xdy)dt,\\
\end{cases}
\end{equation}

\noindent up to terms of order $O((x^{\prime\mu})^2)$
in the coefficients, or even of order $O(\eta^2)$,
since the relation between $x'$ and $\eta$ is
homogeneous.

More precisely,

\begin{equation}\label{E6.22bis}
g^{\prime}_{\mu\nu} = \eta_{\mu\nu} +
O((x^{\prime\mu})^2),
\end{equation}

\noindent where the first order derivatives of
$g^{\prime}$ are vanishing in $P_{\circ}$.

We may also verify that the new coordinate axis are
tangent to the corresponding vectors $e_{\alpha}$ given
in equations (\ref{E6.3}), (\ref{E6.4}) and
(\ref{E6.5}), that is

\begin{equation}\label{E6.23}
e_{\alpha} = \frac{\partial}{\partial x^{\prime
\alpha}},
\end{equation}

\noindent and again with these two conditions satisfied
the transformation (\ref{E6.18}), (\ref{E6.19}) and
(\ref{E6.20}) is unique.

The transformation of the radial and tangential
components of the velocities from $\mathcal S$ to
$\mathcal L^{\prime}$ are

\begin{equation}\label{E6.24}
\dot{x}^{\prime}_{\circ} = \frac{1}{\gamma_{\circ}}
\frac{\dot{r}_{\circ}}
{1-\beta^2_{\circ}-\frac{\beta_{\circ}}{c}r_{\circ}\dot{\phi}_{\circ}},
\end{equation}

\begin{equation}\label{E6.25}
\dot{y}^{\prime}_{\circ} = \frac{r\dot{\phi}_{\circ}}
{1-\beta^2_{\circ}-\frac{\beta_{\circ}}{c}r_{\circ}
\dot{\phi}_{\circ}},
\end{equation}

\noindent where $\dot{r}=\frac{dr}{dt}$ etc., and where

\begin{equation}\label{E6.26}
\begin{cases}
\dot{r}_{\circ} &= \dot{x}_{\circ}\cos{\phi_{\circ}}+
\dot{y}_{\circ}\sin{\phi_{\circ}},\\
 r_{\circ}\dot{\phi}_{\circ} &=
-\dot{x}_{\circ}\sin{\phi_{\circ}}+
\dot{y}_{\circ}\cos{\phi_{\circ}}.\\
\end{cases}
\end{equation}

The coordinate velocities, radial and tangential, are
given by equations (\ref{E6.12}) and (\ref{E6.13}). If
we substitute in (\ref{E6.24}) and  in (\ref{E6.25})
the values of $\dot{r}$ and $\dot{\phi}$ given there we
get for the radial proper velocity

\begin{equation}\label{E6.27}
\dot{x}^{\prime}_{\circ} = \pm c, \quad
\dot{y}^{\prime}_{\circ} = 0,
\end{equation}

\noindent and for the tangential one

\begin{equation}\label{E6.27''}
\dot{y}^{\prime}_{\circ} = \pm c, \quad
\dot{x}^{\prime}_{\circ} = 0,
\end{equation}

\noindent that is once again , the anisotropy, which
was present in $\mathcal S$, in $\mathcal L^{\prime}$
disappears.

We have verified the absence of anisotropy in the light
propagation in two particular cases, for radial
propagation and separately for tangential propagation.
For the general case we have only to take a linear
combination of the tangential and of the radial
velocities. This can be done, but, for sake of
simplicity, we omit that.

Once we have made a clear distinction between the
coordinates $(t;x,y)$ and the proper coordinates
$(t';x',y')$, we may understand better some puzzling
aspects of the rotating disk.

We may work for instance with the set $(t;x,y)$ of
coordinates and follow two closed trips at constant
time $t$ around the rim of the disk, one clockwise and
the other anti-clockwise. Now, the time $t$ is the same
as that of the laboratory frame as stated by the
(\ref{E6.1}), and even the distance covered is the
same, since the radius is unchanged by this
transformation. But the coordinate velocities, as we
have shown in (\ref{E6.13}), are different, so, as has
been shown in \cite{RPet}, the times necessary for
completing a closed path are different. The Sagnac
effect follows from that.

But we may work with the other set of proper variables
$(t';x',y')$, where we should understand that an
infinite collection of local inertial frames is placed
along a path of constant proper time. Now the speed of
light is isotropic and has its value equal to $c$,
indeed, transforming from one frame to the next, the
synchronization prescription doesn't change. What now
change, as shown in \cite{RTar1}, is the length of the
paths, which are now different. Again the Sagnac effect
follows.

We see that no ambiguity or contradiction arises in
this two pictures of the rotating disk. These are
simply two different way of seeing the system, the
distinction between a coordinate speed and a standard
speed being peculiar to the accelerated systems, or,
better, to the not time-orthogonal systems.

The distinction between coordinate and standard
velocity is also stressed in \cite{RWuc}, where various
examples are discussed from the point of view of
standard coordinates and of other systems of
coordinates.

\section{\label{S7}An application}

What we have seen until now can find a nice application
to the analysis of the Selleri's paradox \cite{RSel},
which, to tell the truth, was already discussed by
Rizzi and Tartaglia \cite{RTar2}. Nevertheless, we will
see that the careful distinction between the two
concepts of velocity allows us to have a more clear
understanding of the issue.

The argument is the following: suppose that, on a
rotating disk, two light rays are emitted in opposite
directions along the rim of a disk, and that, after a
complete tour, they come back to the starting point. We
know that there will be a difference in the arrival
times of the two rays. This is the well known Sagnac
effect.

The disk is supposed to be in a laboratory frame, which
is a inertial frame, free of gravitational effects, and
equipped with Minkowski coordinates. Let us call
$t_{\circ}$ ($T$ in our notations) the time measured in
this frame and $t$ the time measured by a clock at rest
on the disk (proper time). Let the angular speed of
rotation be $\omega$ and $L_{\circ}$ be the
circumference of the disk as seen from the laboratory
frame.

The relation between $t_{\circ}$ and $t$ is given by
the usual relation

\begin{equation}\label{E7.1}
t_{\circ} = t F(v, ...),
\end{equation}

\noindent where the factor $F$ is
$\frac{1}{\sqrt{1-(v/c)^2}}$, but this is not very
important, since it will cancel.

The time intervals for the light rays to complete the
tour, as measured in the laboratory, are

\begin{equation}\label{E7.2}
L_{\circ} - x = c(t_{\circ 2} - t_{\circ 1}),\quad\quad
x = c(t_{\circ 2} - t_{\circ 1}),
\end{equation}

\noindent where $x$ is the arc of which is rotated the
disk during the time $t_{\circ 2} - t_{\circ 1}$, and
the first relation determines the arc covered by the
ray of light, which goes in the direction contrary to
the rotation. Analogous relations holds for the other
ray, and, in conclusion, we can write

\begin{equation}\label{E7.3}
t_{\circ 2} - t_{\circ 1} =
\frac{L_{\circ}}{c(1+\beta)},\quad\quad t_{\circ
3}-t_{\circ 1} =\frac{L_{\circ}}{c(1-\beta)},
\end{equation}

\noindent where $\beta = \omega r/c$.  If we multiply
by $F$ the left-hand side of both equations we get the
analogous relations for the proper times,

\begin{equation}\label{E7.4}
(t_2 - t_1) F = \frac{L_{\circ}}{c(1+\beta)},\quad\quad
(t_3 - t_1) F = \frac{L_{\circ}}{c(1-\beta)}.
\end{equation}

The velocities of the two rays be $c_{-}$, for the
counter-rotating ray, and $c_{+}$ for the co-rotating
one. For them we get

\begin{equation}\label{E7.5}
\frac{1}{c_{-}} = \frac{t_2 - t_1}{L};\quad\quad
\frac{1}{c_{+}} = \frac{t_3 - t_1}{L},
\end{equation}

\noindent where $L$ is the length of the rim, as seen
from the disk. Taking into account equation
(\ref{E7.4}), we get for the ratio of the velocities

\begin{equation}\label{E7.6}
\frac{c_{-}}{c_{+}} = \frac{1 +\beta}{1 - \beta}.
\end{equation}

Observe that the factor $F$ doesn't contribute to this
ratio. This is an important point, since it means that
the ratio is the same if we use inertial times
$t_{\circ}$ or proper times $t$.

The paradox arises in the limit $r \rightarrow\infty$,
$\omega\rightarrow 0$, while keeping constant the
product $v = \omega r$. In the limiting case a small
portion of the disk will become equivalent to an
inertial frame, and the speed of light should be $c$ in
both directions, and the ratio (\ref{E7.6}) should be
1. So there is a contradiction and so we have the
paradox.

Rizzi and Tartaglia \cite{RTar2} observed that the
speed of light on the rim is $c$, as they have
demonstrated in \cite{RTar1}, and so the ratio
\underline{is 1} and no contradiction arises when we
perform the limit $r\rightarrow\infty$.

Let us now follow the point of view developed in the
previous Sections. If we follow step by step the
calculation from equation (\ref{E7.3}) until equation
(\ref{E7.6}), we see that, since the factor $F$ and the
length $L_{\circ}$ cancel out, the velocities in
(\ref{E7.6}) are indeed the \underline{coordinate}
ones, calculated with coordinate times (which by the
way are the same as the inertial ones) and with the
length measured in the inertial frame.

So, for \underline{coordinates velocities}, the result
(\ref{E7.6}) is correct, as shown in Section VI,
equation (\ref{E6.13}).

At the same time, the standard velocities are always
$c$, and the analogous ratio is 1.

On the other hand, the result (\ref{E7.6}) is not
correct for the standard velocities. This can be seen
if we observe that the length $L$, used for calculating
the velocities, cannot be the same for both directions.
This is because we cannot invoke any symmetry argument.
These lengths have been calculated by Rizzi and
Tartaglia in \cite{RTar1}. They are

\begin{equation}\label{E7.7}
L_{\pm} = L_{\circ} \sqrt{\frac{1\pm\beta}{1\mp\beta}},
\end{equation}

\noindent where $L_{\circ}$, as before, is the length
of the rim of the disk, as measured in the laboratory,
and is put in their paper equal to $2\pi r$.

If we use these values in equations (\ref{E7.5}), with
$L_{-}$ and $L_{+}$ replacing $L$ in the first and in
the second equation respectively, we get indeed

\begin{equation}\label{E7.8}
\frac{\tilde{c}_{-}}{\tilde{c}_{+}} = 1,
\end{equation}

\noindent where $\tilde{c}_{-}$  and $\tilde{c}_{+}$
are now the \underline{standard} velocities for the ray
propagating in the opposite direction of the rotation
and in the same direction respectively.

It remains to understand what happens in the limit
$r\rightarrow\infty$. But this can be easily understood
if we introduce a set of coordinates on the disk.

Let us consider the transformation from the laboratory
to the disk $(T,X,Y)\longmapsto (t,x,y)$, where we
label with $(T,X,Y)$ the coordinates of the laboratory
and with $(t,x,y)$ those of the rotating system.

We have, as usual

\begin{equation}\label{E7.9}
\begin{cases}
t &= T,\\
x &= X\cos(\omega T) + Y\sin(\omega T),\\
y &= - X\sin(\omega T) + Y\cos(\omega T).\\
\end{cases}
\end{equation}

If we consider a point $P$ at rest on the disk , with
coordinates $(x, y)$, as seen from the laboratory,
inverting the previous equations we can write

\begin{equation}\label{E7.10}
\begin{cases}
T &= t,\\
X &= x\cos(\omega t) - y\sin(\omega t),\\
Y &= x\sin(\omega t) + y\cos(\omega t).\\
\end{cases}
\end{equation}

In terms of polar coordinates $x=r\cos{\phi}$,
$y=r\sin{\phi}$ we have

\begin{equation}\label{E7.11}
\begin{cases}
X = \frac{v}{\omega}(\cos{(\omega t)}\cos{\phi} -
\sin{(\omega t)}\sin{(\phi)}),\\
Y = \frac{v}{\omega}(\sin{(\omega t)}\cos{(\phi)} +
\cos{(\omega t)}\sin{(\phi)}),\\
\end{cases}
\end{equation}

\noindent where we have expressed the radius r in terms
of $\omega$, $r=v/\omega$, where $v$ is the velocity at
the rim. In the limit $\omega\rightarrow 0$, developing
up to the first order in $\omega$ and substituting in
the expression of the metric

\begin{equation}\label{E7.12}
g = -c^2 dT^2 + dX^2 + dY^2,
\end{equation}

\noindent we obtain

\begin{equation}\label{E7.13}
g = -c^2(1 - \beta^2)dt^2 +r^2 d\phi^2 + 2 v (r d\phi)
dt,
\end{equation}

\noindent where $\beta = v/c$, and the motion of the
point $P$ is along a line orthogonal to the direction
determined by the angle $\phi$. Without developing all
the details of the motion, we may recognize the metric
as that of the Galilean transformation, studied in
Section \ref{S4}, where $rd\phi$ takes the place of
$dx$.

We conclude that the frame is not inertial
\cite{RKla2}, because the transformation is linear, but
not Lorentzian. Indeed, a translational motion is not
necessarily given by a Lorentz transformation.

As a consequence, even in the limit, the coordinate
velocities need not be equal to c, and their ratio can
be different from 1. Indeed it agrees with that of
equation (\ref{E7.6}), as can be seen from the
equations (\ref{E4.6}).

Therefore, in the limit, we get the same ratio as for
finite values of $r$ and the paradox is solved.

\bigskip

\section{\label{S8} Conclusions}

\bigskip

In the preceding Sections we have applied a known
mathematical result concerning metric tensors to the
kinematics of non-inertial systems. Given a generic
system, with a given metric tensor, we may define in
the neighborhood of any point of the system a
coordinate set, the standard coordinates, such that the
transformed metric be Minkowskian. This transformation
is almost unique. In terms of these new coordinates the
speed of light is certainly isotropic and has value c.

On the other hands, the original coordinates defining
the system need not be Minkowskian. The speed of light
measured with these coordinates is in general not
isotropic and doesn't have the value c. If the system
was defined starting from an inertial system, which we
call laboratory system, we may see the reason for this
difference. Indeed, transforming from the laboratory to
the system, we necessarily induce a well definite
synchronisation prescription, not necessarily the
standard one (Einstein synchronization), which is only
dictated by the transformation, as seen in the
discussion in Section \ref{S2}.

We have shown all this in some examples: the Galilean
transformation, the hyperbolic motion and the rotating
disk. In all these cases we have exhibited the previous
transformation and we have shown how, transforming from
one set of coordinates to the other, the possible
anisotropy of the light speed present in terms of the
original coordinates, vanishes.

It is clear that the two concepts of velocity have a
different physical meaning. The coordinate speed
follows from the definition of a set of coordinates
which, in the examples we have seen, are not globally
defined. In particular, in the example of the rotating
disk, the time is global, being the same as that of the
laboratory, but there is a degeneration of the metric,
as seen in (\ref{E6.9}): the coefficient of $dt^2$
vanishes when $R\omega = c$.

On the other hand, proper times and standard lengths
are the times and lengths measured by standard clocks
and rods, that is by an observer at rest in the
non-inertial system. Their importance from an
experimental and practical point of view is obvious and
cannot be neglected. This means that both concepts are
important in order to have a clear understanding of the
argument, as stressed in \cite{RPet}.

\bigskip

\section{\label{S9} Acknowledgment}

\bigskip

The author is indebted to L.Lusanna and P.Sona for many
suggestions, for their useful criticism and for reading
the manuscript

\bigskip

\section{Appendix}

\bigskip

In this Appendix we give the definition of the
anisotropy parameter $\epsilon$, quoted in Section
\ref{S3} and in the following Sections.

In order to define the simultaneity of two events, in a
given frame, we may suppose that an observer $A$, who
is moving in an arbitrary way, has the possibility of
emitting a ray of light, which will be reflected or
re-emitted back to him.

Let us call the event at which this reflection happens
$P_B$. The time $t_B$ of this event is what we want to
define.

The observer has a clock and he can measure the time of
emission and the time at which the ray come back.

Let us call the first time $t_A$ and $t_{A'}$ the
second one.

The time $t_B$ is defined by

\begin{equation}\label{EA.1}
t_B = t_A + \epsilon(t'_A -t_A),
\end{equation}

\noindent where the parameter $\epsilon$ can take any
value in the open interval $(0,1)$.

The equation (\ref{EA.1}) defines the simultaneity of
the event $P_B$ with an event lying on the world-line
of the observer.

The symmetric choice $\epsilon = 1/2$ corresponds to
the Einstein synchronization.

It is shown, for instance in \cite{RCap}, that this
definition of simultaneity is as correct, from the
physical point of view, as the usual one, for any
choice of $\epsilon$.

This choice has an important property: together with
the hypothesis of the isotropy of the one-way speed of
light, it implies the isotropy of space \cite{RCap}.

\end{document}